
\input harvmac

\def\figin{\epsfcheck\figin}\def\figins{\epsfcheck\figins}
\def\epsfcheck{\ifx\epsfbox\UnDeFiNeD
\message{(NO epsf.tex, FIGURES WILL BE IGNORED)}
\gdef\figin##1{\vskip2in}\gdef\figins##1{\hskip.5in}
\else\message{(FIGURES WILL BE INCLUDED)}%
\gdef\figin##1{##1}\gdef\figins##1{##1}\fi}
\def\DefWarn#1{}
\def\figinsert{\goodbreak\midinsert}
\def\ifig#1#2#3{\DefWarn#1\xdef#1{fig.~\the\figno}
\writedef{#1\leftbracket fig.\noexpand~\the\figno}%
\figinsert\figin{\centerline{#3}}\medskip\centerline{\vbox{\baselineskip12pt
\centerline{\footnotefont{\bf Fig.~\the\figno:} #2}}}
\bigskip\endinsert\global\advance\figno by1}

\def\Title#1#2{\rightline{#1}\ifx\answ\bigans\nopagenumbers\pageno0
\vskip0.5in
\else\pageno1\vskip.5in\fi \centerline{\titlefont #2}\vskip .3in}

\font\caps=cmcsc10

\noblackbox
\parskip=1.5mm


\def\npb#1#2#3{{\it Nucl. Phys.} {\bf B#1} (#2) #3 }
\def\plb#1#2#3{{\it Phys. Lett.} {\bf B#1} (#2) #3 }
\def\prd#1#2#3{{\it Phys. Rev. } {\bf D#1} (#2) #3 }
\def\prl#1#2#3{{\it Phys. Rev. Lett.} {\bf #1} (#2) #3 }

\def\cmp#1#2#3{{\it Commun. Math. Phys.} {\bf #1} (#2) #3 }

\def\bb#1{{\tt hep-th/#1}}


\def\dj{\hbox{d\kern-0.347em \vrule width 0.3em height 1.252ex depth
-1.21ex \kern 0.051em}}

\def\ee{{\rm e}\,}

\def\Tr{{\rm Tr\,}}

\def\ket{\rangle}
\def\bra{\langle}



\lref\rcom{L. Susskind, L. Thorlacius and J. Uglum, \prd
{48}{1993}{3743.}}
\lref\rgib{G.W. Gibbons and M. Perry, {\it Proc. Roy. Soc. London}
{\bf A 358} (1978), 467.}
\lref\rgidd{S.B. Giddings, {\it ``Comments on Information Loss and
Remnants"}, Santa Barbara preprint UCSBTH-93-35, \bb{9310101.}}
\lref\rhow{S.W. Hawking, \cmp {43}{1975}{199.}}
\lref\rcall{C. Callan, S. Giddings, J. Harvey, and A. Strominger, \prd
{45}{1992}{1005.}}
\lref\rtho{G. 't  Hooft, \npb {256} {1985}{727.}}
\lref\rthof{C.R. Stephens, G. t'Hooft and B.F. Whiting, {\it ``Black Hole
Evaporation without Information Loss"} Utrecht preprint THU-93/20,
hep-gc/ 9305008.}
\lref\rsvv{K. Schoutens, E. Verlinde and H. Verlinde, {\it ``Black Hole
Evaporation and Quantum Gravity".} CERN-Princeton preprint.
CERN-TH.7142/94, PUPT-1441 \bb{9401081.}}
\lref\rbom{L. Bombelli, R.K. Koul, J. Lee and R.D. Sorkin, \prd {34}
{1986} {373\semi}
M. Srednicki, \prl {71} {1993} {666\semi}
C. Callan and F. Wilczek, {\it ``On Geometric Entropy"}, preprint
IASSNS-HEP-93/87 \bb{9401072.}}
\lref\rstrat{D. Kabat and M.J. Strassler, {\it ``A comment on Entropy and
Area"},
Rutgers preprint RU-94-10, \bb{9401125.}}
\lref\rsusk{L. Susskind, {\it ``Strings, black holes and Lorentz
contraction"}, Stanford preprint SU-ITP-93-21, \bb{9308139}\semi
L. Susskind, {\it ``Some Speculations about Black Hole Entropy in
String Theory"}, Rutgers preprint, RU-93-44 \bb{9309145}\semi
L. Susskind and J. Uglum, {\it ``Black Hole Entropy in Canonical Quantum
Gravity
and Superstring Theory"}, Stanford preprint. SU-ITP-94-1 \bb{9401070.}}
\lref\rbann{M. Ba{\~n}ados, D. Teitelboim and J. Zanelli, \prl
{69}{1992}{1849\semi}
G.T. Horowitz and D.L. Welch, \prl {71}{1993}{328.} }
\lref\rpol{J. Polchinski, \cmp {104}{1986}{37\semi}
K.H. O'Brien and C.-I Tan, \prd {36}{1987}{1184\semi}
E. Alvarez and M. Osorio, \npb {304}{1988}{327.} }
\lref\rcasher{A. Casher, E.G. Floratos and N.C. Tsamis, \plb
{199}{1987}{377.} }
\lref\real{E. Alvarez and M. Osorio, \prd {36}{1987}{1175.}}
\lref\rgro{D. Gross, M. Perry and L. Yaffe, \prd {25}{1982}{330.}}
\lref\rgrokleb{D. Gross and I. Klebanov, \npb {344}{1990}{475.}}
\lref\rkog{Ya. I. Kogan, JETP Lett. {\bf 45} (1987) 709\semi
B. Sathiapalan, \prd {35}{1987}{3277.}}
\lref\raw{J. Atick and E. Witten, \npb {310}{1988}{291.}}
\lref\rdeve{H.J. de Vega and N. Sanchez, \npb {299}{1988}{818.}}
\lref\rwitt{E. Witten, \prd {44}{1991}{314.}}
\lref\rDVV{R. Dijkgraaf, E. Verlinde and H. Verlinde,
\npb {371} {1992} {269.}}
\lref\rbar{J.L.F. Barb\'on, {\it ``Horizon divergences of Fields and
Strings in Black Hole backgrounds"}, Princeton preprint, PUPT-1448
\bb{9402004} (to appear in Phys. Rev. D.)}
\lref\rgaw{K. Gawedzki, {\it``Non-compact WZW Conformal Field
Theories"}, Cargese lectures 1991.}
\lref\rbars{I. Bars, {\it ``Superstrings on curved spacetimes"},
USC-92 / HEP-B5. \bb{9210079}. Erice lectures 1992.}


\line{\hfill PUPT-94-1478}
\line{\hfill {\tt hep-th/9406209}}
\vskip 1cm

\Title{\vbox{\baselineskip 12pt\hbox{}
 }}
{\vbox {\centerline{Remarks on Thermal Strings}
\medskip
\centerline{outside Black Holes }  }}

\centerline{$\quad$ {\caps J. L. F. Barb\'on}}
\smallskip
\centerline{{\sl Joseph Henry Laboratories}}
\centerline{{\sl Princeton University}}
\centerline{{\sl Princeton, NJ 08544, U.S.A.}}
\centerline{{\tt barbon@puhep1.princeton.edu}}
\vskip 0.4in
  We discuss the semiclassical approximation to  the level density  of
(super) strings propagating in non-compact coset manifolds $G/H$. We
show that the WKB ansatz agrees with heuristic red-shift arguments
with respect to the ``exact" sigma-model metric, up to some deviations
from minimal coupling, parametrized by the dilaton. This approximation
is used to study thermal ensembles of free strings in black holes,
with the ``brick wall" horizon regularization of 't Hooft.
 In two dimensions the entropy
diverges logarithmically with the horizon thickness, and a local
Hagedorn transition occurs in higher dimensional models.
We also observe that supersymmetry improves the regularity of strings
at the horizon.


\Date{6/94}

\newsec{Introduction}

  Some interesting proposals have been put forward recently about the
role of String Theory in the understanding of black hole
thermodynamics and the information problem \refs\rsusk .
Traditionally, it was assumed that string dynamics was essential only
near space-time singularities. However, as  stressed by 't
Hooft, planckian physics could be required to understand
 horizons, and this clearly opens the applications of
String Theory as a particular model of ultra-short dynamics.
{}From a technical point of view, perhaps the simplest
  problem one can attemp is that of the one loop
entropy, which physically corresponds to the total entropy of the
Hawking radiation by free fields, and appears as the first quantum
correction to the Hawking-Beckenstein entropy in Euclidean Quantum
Gravity. This quantity is known to be
divergent in Field Theory \refs\rtho  ,
\eqn\bek
{ S_{qu} \sim {A_{H}^{(d-2)} \over \epsilon^{d-2}} \left({T\over
T_H}\right)^{d-1}}
where $A_H$ is the horizon area, $\epsilon$ is a short proper-distance
cutoff, and $T_H$ is the Hawking temperature (in two dimensions there
is a logarithmic divergence). It is commonly assumed that this
divergence is equivalent to the information problem, since it leads to
an indefinite storage of information at the horizon. The purpose of a
short distance analysis (i.e. strings) is to provide a physical
definition of the cutoff $\epsilon$.
 There are at least three different derivations of \bek , which
highlight on different aspects of the possible string generalization.

(i) The most basic definition regards $S_{qu}$ as the so called
``geometric entropy" \refs\rbom . The entropy comes from the mixed
state obtained by tracing over local degrees of freedom in a certain
spatial region, and divergences appear associated to the sharp character
of the boundary. In the black hole case one traces over modes with
support on the horizon at late times. In String Theory, it is
intuitively clear that no sharp distinction exists between ``inside"
and ``outside" at the Planck scale, due to the finite size of the
string. However, a rigorous construction of the density matrix as a
trace over configuration space is extremelly difficult, because fixed
loops in the target space are off-shell states in String Theory.

(ii) In the euclidean formal construction we derive $S_{qu}$ from the
euclidean free energy, which in turn is computed as a path integral
over the euclidean section of the black hole, with periodic
identification modulo $\beta$. There is a conical singularity at the
horizon with deficit angle $\beta - 2\pi$, responsible for the
divergence in this formalism.
In String Theory we are instructed to compute the torus path integral
with the euclidean section as target space. Although exact string
backgrounds with black hole interpretation are known, they are based
on non-compact conformal field theories, which makes explicit
computations rather difficult. An expression for the free energy of
the two dimensional string black hole at
$\beta = 2\pi$ was proposed in ref. \refs\rgaw . However, it is very
unclear how to include the conical singularity while keeping conformal
invariance and calculability.

(iii) In a more elementary derivation, one may $\it assume$ a thermal
ensemble as results, for example, from Hartle-Hawking boundary
conditions, and directly compute the partition sum. A reflecting
boundary condition at the horizon is used to cut off the rising local
temperatures, $T/\sqrt{-g_{00}}$ (in momentum space the horizon
divergence appears disguised as an infrared divergence \refs\rbar).
This is the so-called ``brick wall" model of 't Hooft \refs\rtho . Clearly, the
Dirichlet boundary condition at the horizon is rather naive from the
physical point of view. However, it is very appropriate for counting
states,  because it leads to a well defined thermal ensemble.
In the brick wall model, equation \bek\ follows easily from local
red-shift arguments. In String Theory such heuristic
 arguments are suspicious
because of the ``stretching effect" of \refs\rsusk , which suggest
highly non-local string configurations near the horizon, depending on
the frame.
	 One of the results of this
work is to provide a formal discussion of the red-shift arguments.

In this letter we explore the string analog of (iii), which is the
simplest of the three methods (i)-(iii) (It is important to keep in
mind though that nothing ensures $a$ $priori$ the equivalence of the
three approaches in the string case). The WKB
approximation used in \refs\rtho\ for the case of a scalar field in
the Schwarzschild geometry is generalized to strings in a large class
of backgrounds.
 This is the subject of the next
section. In the second and third sections we apply this approximation
to ``stringy" black holes.

\newsec{WKB estimates of String Partition sums}

Consider the canonical partition function of a free string gas:
$$
{\cal Z}(\beta) = \Tr_{\cal H}\,\, \ee^{-\beta :H_0:}
$$
where $\cal H$ is the second quantized Hilbert space built over the
tree level string spectrum (BRST invariant states), $:H_0:$ is the free
Hamiltonian in this space, and the normal ordering means that the
(finite) vacuum energy has been subtracted. Standard Fock algebra
leads to:
\eqn\an
{-{\rm log} {\cal Z} (\beta) = \sum_f (-)^F \Tr_{{\cal H}_f}
\,\,{\rm log}
 \left(
1-(-)^F \ee^{-\beta \omega_f}\right) }
here $f$ labels the different fields (= physical states)
in the string spectrum and ${\cal
H}_f$ stands for the first quantized Hilbert space of each field
(space-time degrees of freedom). The
bosonic or fermionic statistics is incorporated by the sign $(-)^F =
\pm 1$. This representation of the one loop thermal free energy is
general for any background with a timelike isometry,
 and is usually termed as the ``analog
model" because,  to this order (no interactions), the string is
equivalent to the collection of fields in its spectrum.
Using standard zeta function identities,   the integral
representation of the logarithm, and Poisson resummation,
 one can easily derive from \an\
a proper time
representation of the free energy as:
\eqn\ptime
{\beta F(\beta) =
-{\rm log}{\cal Z}(\beta) = \beta  \int_{0}^{\infty} {ds \over
s^2}\,\sum_f \Lambda_f (s)   \sum_{n\neq 0}
(-)^{nF_f} \,\, \ee^{-{\beta^2 n^2 \over 2\pi s}}}
with a vacuum energy integrand given by the general formula:
\eqn\vac
{\Lambda_f (s) = {(-)^{F_f +1} \over 2\pi} \sqrt{s\over 2}
\,\, \Tr_{{\cal
H}_f}\,\, \ee^{-{\pi \over 2} s \omega_f^2} }

This form of the free energy is appropriate for deriving modular
invariant expressions which agree, in all solved flat backgrounds,
with the genus
one path integral computation \refs\rpol . In this case, the proper
time integral is restricted to the fundamental domain of the genus one
modular group, thus avoiding the ultraviolet region (this different
integration domain transforms into \ptime\ after summing over the thermal
winding modes) . Regarding these
manipulations, it is important to keep in mind that the vacuum energy
has been subtracted (there is no $n=0$ term in \ptime ).
  $only$ the renormalized free energy of the
string  equals the sum of the field free energies
(obviously, this is not so for the vacuum energy  since the former
 is U.V. finite while that
of the fields is U.V. divergent). In the brick wall model
it is the renormalized free energy
the one that shows horizon divergences. The equivalence of the analog
model and the path-integral formula for exact black hole backgrounds
is an interesting open problem which we shall not  address here.

 The idea of the semiclassical approximation
is to relate the energies $\omega_f$ to a Schr\"odinger
problem, so that we can evaluate each trace over ${\cal H}_f$ by WKB.
The frequencies $\omega_f$  are determined by the mass-shell
equation of the string ($L_0$ condition). For string models based on
current algebras (gauged WZW models), one can introduce the so-called
``exact" metric and dilaton backgrounds \refs\rDVV, and write the
mass-shell equation as:
\eqn\exact
{\left\{ -{1\over \ee^{-2\Phi}\sqrt{-g}} \partial_{\alpha}
\ee^{-2\Phi}\sqrt{-g} g^{\alpha\beta} \partial_{\beta} + M_f^2
\right\} \Psi_f = 0}
$\Psi_f$ is the vertex operator for the string state $|f\ket$ and the
second order differential operator acts on the zero mode part of
$\Psi_f$. The energies $\omega_f$ are defined by $\partial_t \Psi_f =
-i\omega_f \Psi_f$, once the ``time" coordinate is singled out in the exact
background. $M_f$ is the mass of the state, and it comes entirelly
from the current algebra.

Equation \exact\ is exact for a large class of models. To be as general
as possible, let us consider non-compact cosets of Kazama-Suzuki type
$G/H$. We have a super K\^ac-Moody algebra generated by chiral
currents $j^a (z)$ and fermions $\psi^a (z)$, $a=1,...,{\rm dim}\,\,G$,
with energy-momentum tensor,
$$
T(z) = {1\over k+g} \eta_{ab} :(j^a j^b - \psi^a \partial \psi^b):
$$
where $\eta_{ab}$ has Minkowski signature (one time), and ${\hat k} =
k+g=k+c_2 (G)$ is the level of the corresponding $N=1$ super-WZW
model.
To study the mass-shell condition it is convenient to separate the
zero mode of the chiral Hamiltonian as,
\eqn\elezero
{L_0^G = {1\over k+g} j_0^a j_0^b \,\, \eta_{ab} + N_B^G + N_F^G}
where $N_{B,F}$ are the bosonic/fermionic oscillator number operators:
$$
N_B^G = {2\over k+g}
\sum_{n>0} j_{-n}^a j_n^b \,\,\eta_{ab}
\,\,\,\,\,\,\,\,\,
N_F^G = {2\over k+g} \sum_{r>0} r \psi_{-r}^a \psi_{r}^b \,\,\eta_{ab}
$$
the index $r$ runs over integers in the Ramond sector and
half-integers in the Neveu-Schwartz sector. Due to the relations $[L_0,
j_{-n}^a] = n j_{-n}^a$ and $[L_0, \psi_{-r}^b] = r\psi_{-r}^b$, $N_B$
and $N_F$ are just mode numbers, $\sum n$, $\sum r$ when acting on
Fock space states of the form
$
\prod_j j_{-n_j}^{a_j} \prod_l \psi_{-r_l}^{b_l} | R\ket
$,
where $| R\ket$ is the zero mode or tachyon state. It is labeled by a
representation of the group $G$ (in the Ramond sector, it also carries
the appropriate representation of the $SO({\rm dim} G)$ Clifford
algebra). The zero mode term in \elezero\ acts on the tachyon states
as the quadratic Casimir in the $R$ representation.

The mass-shell condition in the coset model
is obtained by combining left and right
movers,
\eqn\shell
{0 =( L_0 + {\overline L}_0 )^{G/H} -a-{\overline a} = \left[ (L_0 +
{\overline L}_0)^G - (L_0 + {\overline L}_0)^H \right]_{{ \rm zero}
\,\,  {\rm mode}} + M^2} where $a, {\overline a}$ are
                                       normal ordering constants
 (in the standard picture, $a=1$ in the bosonic theory, and $a=1/2 , 0$
     in the NS and R sector respectively). The mass formula,
$$ M^2 = (N + {\overline N})^G -
 (N+{\overline N})^H - (a+{\overline a})$$
 must be supplemented with physical state conditions
(oscillator ``transversality"), plus level matching and GSO
constraints. This definition of mass is natural even if the target
 metric does not approach Minkowski at infinity
 because, under the current rescaling
$j^a \rightarrow \sqrt{k/2} j^a$,
        it goes over the usual definition in the
 flat $k\rightarrow \infty$ limit.
      The important point is that the
 zero mode part of eq. \shell\ can be realized as a second
order differential operator, when acting on the space of functions over
the $G/H$ coset manifold. It is just a linear combination of
Laplacians over the $G$ and $H$ group manifolds with coefficients
depending on the K\^ac-Moody levels (the left and right normalizations
may differ in heterotic models). The functions over the coset are
$H$-invariant: $(j+{\overline j})_H \Psi =0$, and they come labeled
by the left-right zero-mode representantions $(R,{\overline R})$.
Thus, they represent the tachyon part of $\Psi_f$ and we recover
equation \exact . Examples include $G_{-k} /H_{-k}$ bosonic of type
$II$ strings, and heterotic models with bosonic $G_{-k}/H_{-k}  \times
({\rm Gauge})$  right movers and $N=1$ $G_{-k}/H_{-k+g-h}$  left
movers (for a review, see \refs\rbars).

It is convenient to write \exact\  as a Klein-Gordon equation
in the exact metric, upon rescaling $\Psi\rightarrow e^{-\Phi} \Psi$:
\eqn\kg
{\left\{ -\nabla_g^2 + {\widetilde M}_f^2 \right\} \Psi_f \,\,\,
\equiv \,\,\, \left\{ -\nabla_g^2 + M_f^2 +
 (\nabla_g \Phi)^2 - \nabla_g^2 \Phi
\right\} \Psi_f =0}

 In this way we summarize all ``stringy" modifications to minimal
coupling in the effective mass ${\widetilde M}_f$, which includes an
additive renormalization depending on the exact   dilaton.
For black hole applications we are interested in rotationally
invariant situations, where the exact metric is parametrized as:
$$
ds^2 = -\lambda(r) dt^2 + {dr^2 \over \mu(r)} + r^2 d\Omega_{d-2}^2
$$
One can then factorize the angles and rewrite \exact\ as a Schr\"odinger
equation for the frequencies $\omega_f$:
\eqn\sch
{\left\{-{1\over 2}{d^2\over dx^2} + V_{\ell}^f (x) \right\}
 \Psi_{n\ell}(x) =
{\omega_{n,\ell}^2 \over 2} \Psi_{n\ell} (x) }
with the following effective potential:
\eqn\effpot
{V_{\ell}^f = {d-2 \over 4} \gamma(r) \left( {\gamma' (r) \over r} +
{d-4 \over 2r^2} \gamma (r)\right) + {\lambda (r) \over 2} \left(
{C_{\ell} \over r^2} + {\widetilde M}_f^2 \right) }
here $\gamma (r) = \sqrt{\lambda(r) \mu(r)}$, $\ell$ denotes the angular
momenta,  $C_{\ell}$ is the
quadratic Casimir of the $SO(d-1)$ rotation group, and prime means
$d/dr$. The variable $x$ appearing in \sch\ becomes simply a
``tortoise" radial coordinate $x= \int dr/\gamma(r)$.
In the presence of regular horizons, $
\lambda(r) \sim \lambda_0' (r-r_0)$, $\mu(r) \sim \mu_0' (r-r_0)$, and
for non-singular dilaton fields, we have the usual properties
\refs\rbar : the
horizon appears at $x=-\infty$ and the potential vanishes
exponentially there, $V_{\ell} (x) \sim C {\rm exp}(4\pi T_H x)$.
As a result, the positive spectrum $\omega_f^2$ remains continuous even
after enclosing the system in a large box (located at $x_+ \gg 1$),
and a second ``wall" (at $x_- \ll 0$ near the horizon), is needed to
avoid infrared problems. We thus conclude that the string black hole
still allows a continuous spectrum of states as long as the exact
metric has horizons and the dilaton field is well behaved there.

Some comments are in order regarding Dirichlet walls in String Theory.
It is known that a Dirichlet condition on the full coordinate field
$X^\mu (z,{\bar z})$ leads to a divergent vacuum energy \refs\rcasher
. This is not of direct relevance here because we are conventionally
subtracting vacuum energies in discussing thermodynamics. On the other
hand, $X^{\mu} (z,{\bar z})$ is not a $bona$ $fide$ physical operator
and it is not surprising that pathologies appear. Instead, we impose
Dirichlet conditions on the equation \exact . The result is a certain
projection (discretization) of the spectrum of zero modes, similar to
the case of toroidal compactifications, although more complicated. In
this way we preserve the conformal structure and world-sheet current
algebra.

So far all formulas were exact, but the WKB approximation proceeds
 from \sch\ by simply using
 the semiclassical formula for the level density when computing the
space-time trace:
$$
\Tr_{{\cal H}_f} \rightarrow {1\over\pi} \sum_{\ell} \int d\left(
\int \,dx\,\sqrt{\omega^2 - 2V_{\ell}^f (x)}\right)
$$
For example, the result for the vacuum energy integrand per field degree
of freedom \vac\ reads:
$$
\Lambda_f^{WKB} (s) = {(-)^{F_f +1} \over (2\pi)^2} \sum_{\ell}
\int_{x_-}^{x_+} dx \,\,\ee^{-s\pi V_{\ell}^f (x)}  $$
In this form the infrared problems become obvious when the effective
potential vanishes in any of the limits $x_{\pm} \rightarrow
\pm\infty$, and this occurs at the horizon for $any$ mass
 (in terms of the vertex operators $\Psi_f$, they
oscillate wildly as the target fields approach the horizon).
Since $x$ is interpreted as a radial coordinate, we see that the WKB
method naturally produces integrated densities over space. To be more
explicit, if we further approximate the angular momentum sum by a smooth
integral we obtain
\eqn\last
{\Lambda_f^{WKB} (s) = {(-)^{F_f +1}\over \pi^d 2^{1+d/2}}
  \int \, d({\rm Vol})
\,\,(\sqrt{\lambda})^{1-d} \,\, s^{1-{d\over 2}} \,\,\ee^{-s\pi V_0^f}
}
where  $d({\rm Vol})$
 is the spatial volume
element and $V_0^f$ denotes the s-wave effective potential that
results form \effpot\ by putting $C_{\ell} =0$. This expression
resembles the flat space one with a position dependent ``mass"
given by $\sqrt{2 V_0^f}$. The red-shift factor $(\sqrt{\lambda})^{1-d}$,
when combined with
 the appropriate power of $
\beta$,  yields the natural scaling on the red-shifted temperature,
and also blows up at the horizon. So, it seems that the horizon
instabilities add up for the different fields in the string spectrum.
In the next section we argue that this is not exactly true.

\newsec{Local Hagedorn singularity}

According to equation \kg , red-shift arguments in the exact
background       should give good
estimates of the thermodynamic densities, up to the dilaton terms that
deviate from minimal coupling to the metric.
It  is easy to see that WKB
is the formal counterpart of  red-shift arguments.  By
directly plugging the WKB ansatz in \an\  one obtains after some
manipulations,
\eqn\ann
{\beta F^{WKB} = -{\Gamma(d/2) \over \pi^{d/2} \Gamma(d)}\int d({\rm Vol})
\,\,  {\tilde
\beta}^{1-d} \sum_f \int_{\beta\sqrt{2V_0^f}}^{\infty} {dz (z^2 -2\beta^2
V_0^f)^{d-1 \over 2} \over \ee^z - (-)^{F_f}} }
where we  have estimated
again the discrete sum over angular momenta $(\ell)$ by a smooth
integral. We see that, up to
the back scattering effects (first term in formula \effpot ), all
local dependence goes in the combination ${\tilde \beta} {\widetilde
M}_f$, for ${\tilde \beta}=\beta\sqrt{\lambda}$
 the red-shifted inverse temperature. Each term in \ann\ has now the
form of the flat space free energy density of free bosons/fermions
with a local red-shifted temperature and a local mass. Formally, the
WKB approximation is good as long as the effective potential \effpot\
is a smooth function ( it is exact in flat backgrounds).
However, for a fixed brick-wall cutoff $x_-$ , the WKB approximation
gets worse as we consider higher masses or angular momenta
in the sum \ann , because the
effective potential becomes rather steep near $x_-$. Nevertheless,
the WKB method gives correctly the leading exponential
supression at large masses, since this follows basically from scaling
and dimensional analysis (for a discussion of these issues see
\refs\rstrat ).

 Equation \ann\ is very appropriate for a large mass expansion.
Assuming that the dilaton field is regular and bounded, and defining
the density of levels $\rho(M) = \sum_f \delta (M-M_f)$ we have the
asymptotic estimate,
$$
\beta F^{WKB}  \rightarrow -{\Gamma(d/2) \over \pi^{d/2}}
\int d({\rm Vol}) \,\,{\tilde \beta}^{1-d} \sum_M \rho(M)
\,\,\ee^{-{\tilde
\beta} M}  + {\cal O} \left({1\over M}\right)  $$
(as previously stated, the constants in front may be wrong, but this
will not change the conclusion).
                On general grounds,
 we expect the density of states to diverge exponentially for
string theories above two dimensions, $\rho (M) \sim {\rm
exp}(M\beta_{\rm Hag})$ . This is a physical requirement
because the string spectrum of any black hole solution should approach
the flat space spectrum in the asymptotically flat region.
As a result,  there is a local Hagedorn transition
where the local red-shifted temperature reaches the critical value
(the same singularity follows form \ptime and \last\ as a pole at $s=
\infty$).
 This means that we are unable to discuss the horizon physics of
``stringy" black holes in interesting dimensions, unless we understand
first the old Hagedorn problem.

An interesting example is Rindler space, which corresponds formally to
extremelly massive black holes. We can apply the formalism of the
previous section because the world-sheet conformal field theory is
related to the Minkowkian one by a field redefinition. This means that
current algebra dependent quantities like the mass formula stay the
same and the mass-shell equation is given by \exact\ with
$g_{\alpha\beta}$ the Rindler metric and $\Phi =0$. The associated
Schr\"odinger problem \sch\ becomes a Bessel equation with effective
potential $V^f (x) = (k_{\perp}^2 + M_{f}^2 )\lambda(x)/2$, where
$\lambda(x) = e^{4\pi T_H x}$ and $k_{\perp}$ is the transverse
momentum. The horizon regularization is taken again as a reflecting
wall, which discretizes the frequency spectrum (note that our horizon
cutoff differs from the one in \refs\rdeve , where a deformation in
the Rindler mapping was used instead, resulting in a deformed mass
formula). Integrating the transverse momenta we get for the free
energy:
$$
\beta F^{WKB}_{\rm Rin.} = -\int_{0}^{\infty}{ds\over s^{1+d/2}}
\sum_f \ee^{-{s\over 2}M_f^2}\int d({\rm Vol}){\tilde \beta} \sum_{n\neq
0} {(-)^{(n+1)F_f} \over 2(2\pi)^{d/2}} \ee^{-{{\tilde \beta}^2 n^2\over
2 s}}
$$
This equals the Minkowski formula with $\beta$ replaced by the local
red-shifted inverse temperature ${\tilde \beta}(x) =
\beta\sqrt{\lambda(x)}$. Since $M_f$ is the Minkowskian mass,
 a modular invariant
expression follows easily using the tricks in \refs\rpol . Focusing on
the bosonic case to simplify notation, the WKB one-loop free energy of
strings in Rindler space takes the form
\eqn\guapa
{\beta F^{WKB} = \int_{\cal F} {d^2 \tau \over ({\rm Im} \tau)^2}
\Lambda (\tau) \int d({\rm Vol}) \left( \Theta_{\tilde \beta} (\tau)
-1\right) {\tilde \beta} }
Here $\Lambda (\tau)$ is the flat space modular invariant cosmological
constant integrand, $\cal F$ is the genus one fundamental region of
the modular group, and $\Theta_{\tilde \beta} (\tau)$ is the well known
thermal theta function, coming from the winding modes of the torus
cycles onto the thermal circle of length $\beta \sqrt{\lambda(x)}$ at
each point:
$$
\Theta_{\tilde\beta} (\tau) = \sum_{m,n=-\infty}^{+\infty}
\ee^{-{{\tilde \beta}^2 \over 2\pi {\rm Im}\tau} | m\tau + n|^2}
$$
Possible extra compactified dimensions are already taken into account
in $\Lambda (\tau)$, and supersymmetric generalizations are
straightforward by changing $\beta\rightarrow {\tilde \beta}$ in the
Minkowskian densities. Formula \guapa\ is one of the main results of
this paper. In shows that modular invariant, U.V. finite expressions
can be derived in Rindler space in a well defined approximation. It
also shows the Hagedorn problem as the space integral approaches the
horizon, although in this case we can speculate about the high
temperature phase using the duality properties of the theta function.
The free energy density close to the horizon $({\tilde\beta}\rightarrow
0)$ behaves as
$$
\beta f(x\rightarrow -\infty) \rightarrow {\rm const.} \times
\,{\tilde\beta}^{-1} \int_{\cal
F} {d^2 \tau \over ({\rm Im}\tau)^2} \Lambda (\tau)
$$
We see the typical two-dimensional Atick-Witten form with a
coefficient given by the vacuum energy density of the string in flat
space. For bosonic strings this is I.R. divergent due to the zero
temperature tachyon, and for supersymmetric strings we expect the
cosmological constant to vanish. As a result, perturbative String
Theory is regular at the horizon precisely in the supersymmetric case,
assuming of course that the high temperature phase given by duality
has any physical significance (which is questionable, for example, the
canonical entropy is negative in this phase). This behavior is
specific of the duality properties of String Theory, since
supersymmetric field theories are still ill defined at the horizon:
using the identity $\beta F_F (\beta) = \beta F_B (\beta) - 2\beta F_B
(2\beta)$ between the fermionic/bosonic free energies per degree of
freedom, and the asymptotics in \ann , it is easy to see that the
horizon divergence proportional to $\int d({\rm Vol})
{\tilde\beta}^{1-d}$ never cancels between a finite number of bosons and
fermions.

\newsec{Two dimensional black hole}
                The $SL(2,R)/U(1)$ black hole \refs\rwitt\ is an
interesting example for many reasons. The only propagating degree of
freedom is a single tachyon mode (discrete states are global modes
and
we do not expect them to  drive horizon divergences). In addition,
the WKB approximation is reliable because it only affects the massless
tachyon, and there is no Hagedorn transition.
 The interesting question to answer is whether the ``stringy"
dilaton effects of \kg\ manage to smooth out the horizon
instability.
 The exact metric and dilaton fields read \refs\rDVV :
$$
ds^2 = {k-2 \over 2} (dr^2 -\gamma^2 (r) dt^2)
$$
$$\Phi (r) = {1\over 2} {\rm log} {\gamma (r)\over {\rm sinh}\, r}
\,\,\,\,,\,\,\,\,\,\,\,\,\,\gamma(r) = 2\left( {\rm coth}^2 {r\over 2} -
{2\over k} \right)^{-1/2}
$$
Tachyon vertex operators are given in terms of hypergeometric
functions. In the axially gauged model they can be expressed as matrix
elements
   $
\Psi_{\lambda, \omega}(r,t) = \bra \ell,\omega | g(r,t)|\ell,
-\omega\ket
$
  in the continuous representations of $SL(2,R)$, $\ell = -1/2 +
i\lambda$. The mass-shell condition $L_0 =1$ leads to $\omega^2
=9\lambda^2$ at the critical value $k=9/4$, and $\lambda$ may be
interpreted as a radial momentum. After the appropriate rescalings,
 the differential representation of
the mass-shell equation is,
$$
\left\{ -{1\over 2}{d^2\over dx^2} + {k-2 \over 2} \gamma^2 (x)\Bigl[ (\nabla
\Phi)^2 - \nabla^2 \Phi -2\Bigr]\right\} \Psi_{\omega} = 2\omega^2
\Psi_{\omega}
$$
where we have changed variables to tortoise coordinates $x=\int
dr/\gamma (r)$ and $\omega$ differs from the previous sections by a
factor of $2$ (note that this eigenvalue problem is not the same as
those in \refs\rDVV ).
The effective potential is completely determined by dilaton effects,
and has the form:

$$
V_{\rm eff} (x) = {\gamma^2 (x) \over 4k} \left( {1\over {\rm sinh}^2
r}{\gamma^2 (x) \over 4k}\left(1+{\gamma^2 (x)\over 4k}\right) +{k-2\over
8} R(x)\right)
$$
here $R(x)$ is the scalar curvature (the only remaining term in the
``low energy" $k\rightarrow \infty$ limit). Again, this potential is regular
at the horizon and vanishes exponentially with the tortoise coordinate
$V_{\rm eff} \sim C\ee^{4\pi T_H x}$. Thus, the dilaton fails
to provide a ``wall" at the horizon and the $x_-$ cutoff must be
imposed by hand in order to prevent infrared problems. This is
achieved by setting
   $
\Psi_{\omega_n} (r_{\pm}, t) = 0 $,
which produces the desired discretization of the spectrum. An
interesting  point is that a generic choice of the inner and
outer walls $r_{\pm}$ projects out the discrete states, because in
general $\omega_n (r_{\pm})$  falls out of the discrete states
list. This is a very natural phenomenon, which occurs in the $c=1$
compactified model  as well (at non-rational multiples of the self-dual radius
$R=1/\sqrt{2}$ the spectrum reduces to tachyons and the zero momentum
discrete states such as $\partial X {\overline \partial}X$).

The WKB analysis along the lines of the previous sections is now
straightforward. The free energy density near the horizon $(x\ll 0)$
is given by
$$
\beta f(x) \simeq -{\pi \over 6} {1\over \beta \gamma (x)}
$$
and the entropy diverges linearly with the ``tortoise distance" to the
horizon: $S\sim \pi \Delta x_- /3\beta$. At the Hawking
temperature we get the well known universal result
\eqn\univ
{S_{\rm boundary} = {1\over 6}\,\, {\rm log}\,\,  {1\over\epsilon}}
where $\epsilon$ is now a proper distance cutoff.
In general, as long as the dilaton is regular in the horizon region,
the local geometry is still asymptotically of Rindler type, and we
will get \univ . However, in some situations the dilaton effects may be
important. For example, if we consider the canonical ensemble
defined in the dual background (region V of the black hole), we simply
replace $\gamma (r) \rightarrow {\tilde \gamma} (r) $ which is
obtained by the substitution ${\rm coth} \,{r\over 2} \rightarrow {\rm
tanh} \, {r\over 2}$. The red-shifted local temperature tends to zero
at the singularity $r_0 =2\, {\rm tanh}^{-1} \sqrt{2/k}$, but the
dilaton contribution is singular and negative, and produces an
effective potential unbounded from bellow.

\newsec{Conclusions}

We have used semiclassical methods to study some thermal properties of
free strings in the ``brick-wall" model of 't Hooft. We find that, in
this approximation, ``stringy" effects amount to work in the exact
sigma-model metric, with a local mass renormalization due to the
dilaton. The perturbative spectrum in the horizon region is still
continuous like in Field Theory, but the
 horizon physics is strongly dimension-dependent.
For  higher dimensional models the horizon
problem is shadowed by the  the Hagedorn problem and the stretched
horizon starts abruptly at the Hagedorn radius where
$T/\sqrt{-g_{00}} = T_{\rm Hag}$. It is interesting to note that a
``soft" Atick-Witten high temperature phase \refs\raw : $\beta f(r)
\sim T/\sqrt{-g_{00}}$, leads to a logarithmic divergence of the type
\univ\ , unless the coefficient in front vanishes, as is the case for
supersymmetric strings. Generally
speaking, the Hagedorn transition opens the possibility for new string
effects beyond the one-loop thermal physics, but also puts them beyond
the reach of elementary methods. It
is doubtfull that a perturbative treatment based on a fixed background
will suffice. This conclusion readily applies to the two-dimensional
case, where the absence of Hagedorn singularity is compatible with the
idea that there is no ``stretching effect" for two-dimensional
strings. In fact, the target space effective theory is renormalizable
in two dimensions and should provide a pure field-theoretical
resolution of the logarithmic singularity that appears in this case.

It would be very interesting to have        a path
integral computation (along the lines of point (ii) in the
introduction), with an euclidean understanding of the local Hagedorn
transition. If, on the other hand, such calculation gives a finite
 result,
one should conclude that the
``analog model" and the euclidean path integral
are no longer equivalent definitions of the string thermodynamics in
curved backgrounds (this possibility was suggested in ref.
\refs\rbar\ on the basis of the first quantized path integral
representations in field theory).

\newsec{Acknowledgements}
I am indebted to I. Klebanov, I. Kogan and H. Verlinde for discussions.
This work was supported by NSF PHY90-21984 grant.

\listrefs
\bye